%% file: ijcai24.tex
\documentclass{article}
\usepackage{ijcai24}

\usepackage{times}
\usepackage{soul}
\usepackage{url}
\usepackage[hidelinks]{hyperref}
\usepackage[utf8]{inputenc}
\usepackage[small]{caption}
\usepackage{graphicx}
\usepackage{amsmath}
\usepackage{amsthm}
\usepackage{booktabs}
\usepackage{algorithm}
\usepackage{algorithmic}
\usepackage[switch]{lineno}

\usepackage{tabularx}
\usepackage{multirow}

\title{Enabling Explainable Recommendation in E-commerce with LLM-powered Product Knowledge Graph}

\author{
Menghan Wang 
\and
Yuchen Guo \and
Duanfeng Zhang \and
Jianian Jin \and
Minnie Li \and
Dan Schonfeld \and
Shawn Zhou 
\\
\affiliations
eBay Inc.\\ 
\emails
\{menghawang, yuchguo, duzhang, jianijin, mli7, dschonfeld, xzhou6\}@ebay.com 
} 

\begin{document}

\maketitle

\newcommand{\method}{LLM-PKG}

\begin{abstract}
    How to leverage large language model's superior capability in e-commerce recommendation has been a hot topic. In this paper, we propose \method{}, an efficient approach that distills the knowledge of LLMs into product knowledge graph (PKG) and then applies PKG to provide explainable recommendations. 
Specifically, we first build PKG by feeding curated prompts to LLM, and then map LLM response to real enterprise products. To mitigate the risks associated with LLM hallucination, we employ rigorous evaluation and pruning methods to ensure the reliability and availability of the KG. Through an A/B test conducted on an e-commerce website, we demonstrate the effectiveness of \method{} in driving user engagements and transactions significantly. 
\end{abstract}

\input{introduction}

\input{relatedwork}

\input{approach}

\input{experiment}

\input{conclusion}

\bibliographystyle{named}
\bibliography{references}

\end{document}

%% file: introduction.tex
\section{Introduction}
\label{sec:intro} 

The recent progress in Large Language Models (LLMs) has offered excellent capabilities in understanding, generating, and reasoning with human-like text.  
It is a promising tool to augment recommendation systems by articulating the rationale behind their recommendations in a manner that is comprehensible to users. This not only enhances users' trust and confidence in the recommendations but also fosters a deeper engagement with the recommender system.

Nowadays modern recommender systems on e-commerce websites are complicated and extremely sensitive to the response time, making calling LLM in real-time an unacceptable solution. So a natural idea is to distill useful knowledge from LLM into a knowledge graph (KG) and then apply the KG into recommendation for E-commerce. Trained on vast amounts of textual data and external knowledge sources, LLM is assumed to own world knowledge. So it is a relatively mild condition that LLM understands most usecases of products and user intention behind a product purchase. For instance, carnations are the official flower for mother's day gift. However, it is extremely difficult to mine this kind of relation from data collected by e-commerce website. Customers typically don't express their explicit intention (e.g., they buy carnations for their mothers) during their interactions with the website, which results in a fact that building product knowledge graph is very expensive and labor-intensive in e-commerce. We argue that LLM will narrow this gap significantly and enhance recommendation performance. 

To this end, we propose a new paradigm for product knowledge graph construction with LLM and enterprise data for explainable recommendation. Specifically, we first leverage LLM to generate an initial product knowledge graph (PKG), and then refine the PKG with enterprise data to make it fit more to the target use scenarios. The proposed framework is named \method{} and we provide two kinds of recommendation scenarios: item-based and user-based recommendations. Experiments conducted on an E-commerce website with online traffic showed promising results of our method.

%% file: relatedwork.tex
\section{Related Work}
\label{sec:relatedwork}

Recent advancements in recommendation systems research have seen a surge in methodologies that harness LLMs and knowledge graphs to enhance recommendation effectiveness, transparency, and interpretability. 


1.	\textbf{Utilization of LLMs in Recommendation Systems:} Several studies focus on harnessing the capabilities of LLMs to improve recommendation accuracy and personalization. For instance, \cite{wang2023recmind} and \cite{luo2024kellmrec} explore the use of LLMs for generating personalized recommendations while leveraging additional knowledge sources. Hou et al. \cite{hou2024large} evaluate the zero-sample sorting ability of LLMs in recommendation systems, employing suitable prompt templates combined with conditions, candidates, and sorting instructions. Gao et al. \cite{gao2023chat} propose that LLM can better understand user preferences and build connections between users and items based on in context learning. While these approaches emphasize recommendation effectiveness, they may prioritize accuracy over transparency in explanation generation.

2.	\textbf{Fusion of LLMs with KGs in Recommendation Systems:}  Other works explore the synergy between LLMs and KGs to enhance recommendation tasks. For example, Shu et al. \cite{shu2024knowledge} introduce the KG-LLM Framework, demonstrating its potential for multi-hop link prediction tasks in KGs. Xie et.al \cite{runfeng2023lkpnr} present a framework that combines LLMs and KGs into semantic representations to address long tail problem of inactive users in news recommendation. Furthermore, Abu-Rasheed et.al \cite{abu2024knowledge} utilize KGs as a source of factual context for LLM prompts, mitigating the risk of model hallucinations and ensuring application-intended learning contexts. Although focused on diverse domains, these approaches share common ground with our proposed methodology, highlighting the versatility and effectiveness of leveraging LLMs and KGs in recommendation systems.

In comparison to these existing approaches, our approach offers a comprehensive solution that seamlessly integrates recommendation effectiveness with transparent explanation generation. By combining the strengths of LLMs and KGs, our approach addresses the dual challenges of recommendation accuracy and explainability, ultimately enhancing user trust and satisfaction in recommendation systems.

%% file: approach.tex
\section{LLM-PKG}
\label{sec:approach}

\begin{figure*}[ht]
\centering
\includegraphics[width=1\linewidth]{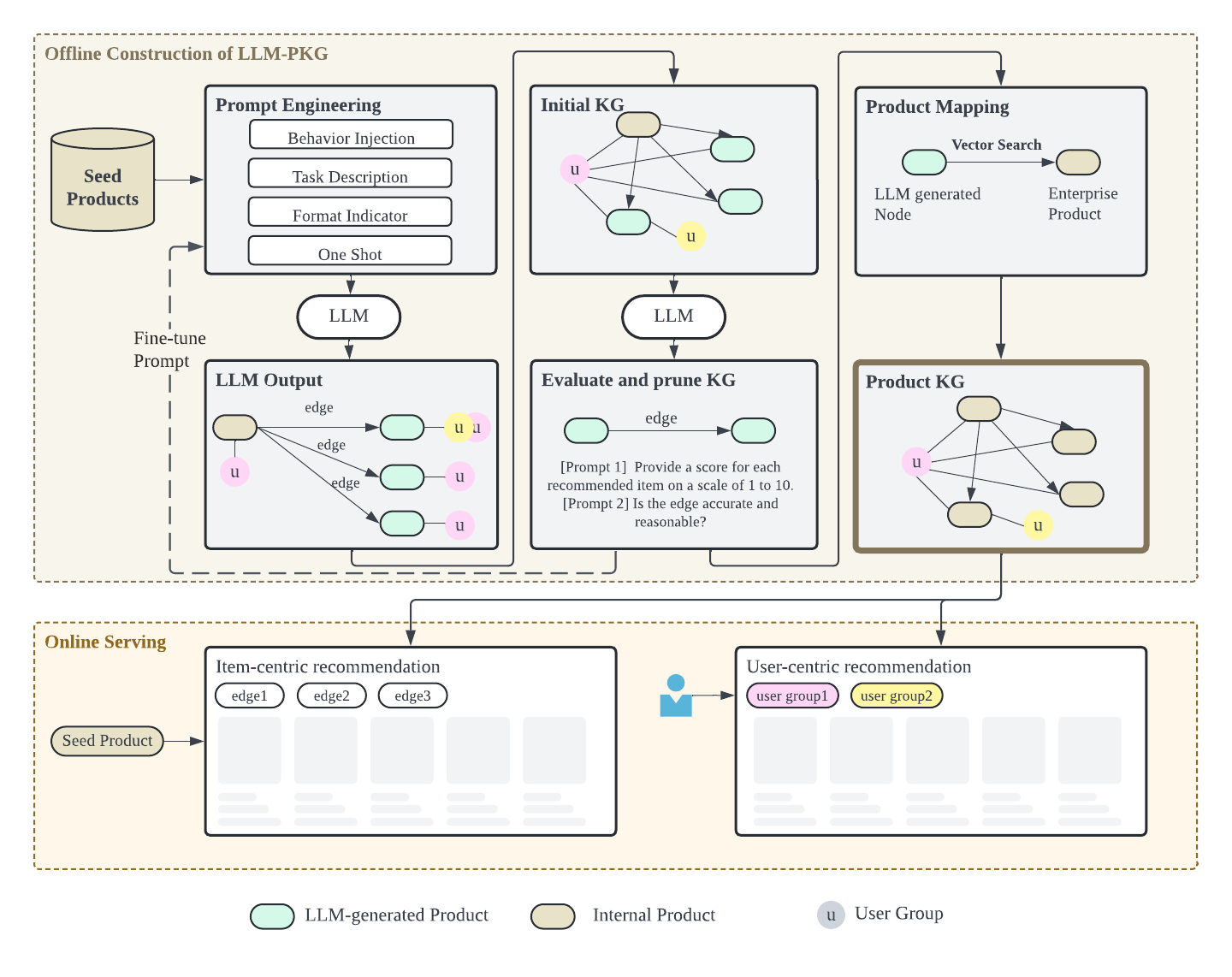}
\caption{\label{fig:framework} Overall Framework}
\end{figure*}

This section provides a detailed introduction of \method{}. We first present the overall structure of \method{} and then elaborate each component in subsequent subsections.

\subsection{Framework Overview}
As shown in Figure \ref{fig:framework}, \method{} consists of two modules: offline construction and online serving. The main processes of offline construction are 1) we employ LLM to infer the recommendations and corresponding rationales between products, then combined the results into an initial knowledge graph which is solely based on the LLM output data. This KG serves as the foundation for capturing implicit relationships and recommendations between products. Subsequently, we leverage LLM to evaluate and refine the KG by pruning existing KG and fine tune the prompts. After iterations we could get a quality KG. Then we map the nodes of the KG to enterprise products/inventory. This mapping facilitates the integration of LLM's comprehensive knowledge and real recommendation use cases. Finally we get the LLM-powered product knowledge graph, which is ready for downstream usecases. In online serving stage, we provide two classical kinds of applying \method{} for recommendation: item-based and user-based recommendation. Pivoted on one node in the KG, we fetch its neighbors as well as connected edges to compose explainable recommendations. Real-world use cases are shown in Sec. \ref{sec:experiment}.

\subsection{Construction of \method{}}
A common usage of LLM is to write prompts to generate desired responses from LLM. So the first step is to design proper contextual prompts to ensure more accurate and contextually relevant results from LLM. 

\subsubsection{Prompt Design}
Inspired by the methodology outlined in \cite{liu2023chatgpt}, we construct prompts with four key components: user behavior injection, task description, format indicator, and an example of input-output pair (one-shot learning). 

The user behavior injection contextualizes the input seed product using natural language, such as, \textit{"A user of an e-commerce website is browsing a product titled \{seed\_product\}"}. This contextualization aims to simulate real user interactions, providing meaningful input to the recommendation model. In practice, the seed product are curated from enterprise product data. We primarily use titles of seed products in experiment though other information like product description could also be used.

The task description specifies the recommendation task, 
 guiding the model on the desired output. For instance, \textit{"Provide \{k\} recommendations that the user might be interested in related to the seed product. For each result, give a reasonable rationale within 5 words, then predict the brand, type, and target audience demographics of each product."} We restrict the rationales to be a phrase within 5 words in order to be better displayed on a user interaction page, such as \textit{'similar style'}, \textit{'affordable alternative'}, etc. 
 
 The format indicator ensures that the output recommendations adhere to a structured format, enhancing their parsability. We instruct the output format to be JSON, facilitating seamless integration of results into a knowledge graph database in subsequent stages. 
 
\begin{figure*}[ht]
\centering
\includegraphics[width=0.9\linewidth]{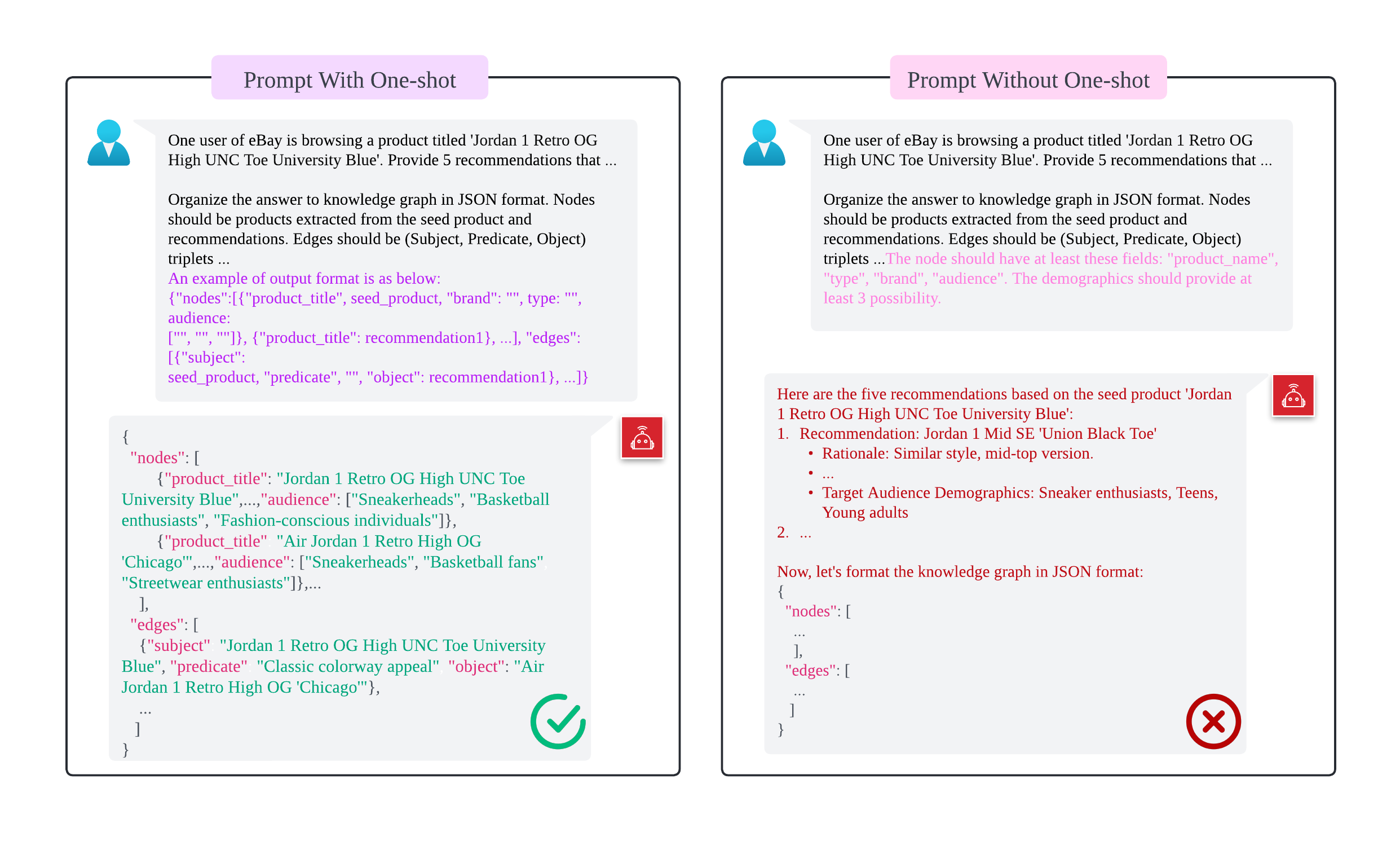}
\caption{\label{fig:oneshot} Effects of One-shot learning.}
\end{figure*}

Additionally, the inclusion of a one-shot learning component significantly improves the correctness rate of the output format. By providing an example of the desired output, the one-shot component guides the model towards generating more accurate results. Specifically, the output of the LLM transforms into a structured JSON string without redundant words. This streamlined format simplifies the post-processing of the outputs. The effect of one-shot learning is illustrated in Figure \ref{fig:oneshot}. The prompt serves as the input to the LLM, guiding it to generate recommendations and corresponding rationales based on a specified seed product. An example template of the prompt used for graph construction is as Table~\ref{table:KG generation prompt}.

\begin{table*}
  \centering
  \begin{tabular}{p{4cm}p{11cm}}
    \toprule
    Component     & Prompt     \\
    \midrule
    User behavior injection & A user of a eBay is browsing a product titled \{seed\_product\}, hereinafter referred to as the seed product. \\
    \hline
    Task description & Provide \{k\} recommendations that the user might be interested in related to the seed product. For each result, offer a reasonable rationale in 5 words, then predict the brand, type, and target audience demographics for each product. \\
    \hline
    Format indicator & Organize the answer to knowledge graph in JSON format. Nodes should be products extracted from the seed product and recommendations. Edges should be (Subject, Predicate, Object) triplets.\\
    \hline
    One-shot example & \{"nodes":[\{"product\_title", seed\_product, "brand": "", type: "", audience: ["", "", ""]\}, \{"product\_title": recommendation1\}, ...], "edges": [\{"subject": seed\_product, "predicate", "similar style", "object": recommendation1\}, ...]\} \\
    \bottomrule
  \end{tabular}
  
   \caption{\label{table:KG generation prompt}Prompt template for \method{} construction}
\end{table*}

\subsubsection{Compose LLM Response into KG}
To simplify the procedure of importing LLM response into KG, we define the format of response to be RDF(Resource Description Framework). RDF is a data format where every piece of knowledge can be broken down into a structure (subject, predicate and object) called triple, which is widely used to represent knowledge graphs. We use products as subject and object, and product relations as predicate. So they are in the form of \textit{(Subject, Predicate, Object}) triplets. Furthermore, nodes representing user groups are derived from user demographics information and added to the graph. Connections between product nodes and the corresponding user group nodes are established to facilitate user-based recommendation. Each product node includes essential attributes such as product name, type and brand. With those steps, we can ensure the comprehensibility of recommendation relationships within the graph structure. After these steps, we can easily import KG into graph databases.


\subsection{Validate and Refine \method{}}

To ensure the quality of the constructed \method{} we require a validation and refinement process. As LLM has excellent \textit{role-play} capabilities we ask LLM to evaluate the quality of nodes and edges in the KG with prompts.

\subsubsection{Validation Method}
Key facets of the edge validation process encompass scoring each recommended product and evaluating the rationale behind recommendations.
\begin{itemize}
\item \textbf{Scoring Recommended Products}: We adopt a scoring mechanism to assess the relative quality of each recommended product. We prompt LLM to assign a score ranging from 1 to 10 to each seed product and recommended product pair based on their acceptability and relevance. This scoring mechanism serves as a quantitative measure of the suitability of recommended items within the context of user preferences and needs. We then use the score to prune nodes. In practice, nodes with score smaller than 6 are removed.  
\item \textbf{Assessing Recommendation Reasons}: Beyond scoring recommended products (nodes in KG), we also use LLM to judge the accuracy and appropriateness of the given recommendation reasons (relationship of the edge). If deemed inaccurate or inappropriate, LLM is prompted to provide a more suitable rationale. This qualitative assessment ensures the alignment of recommendation rationale with user expectations and enhances the interpretability of the recommendation process.
\end{itemize}
An example of validation prompt and LLM output is shown in Table \ref{table:example output} .

\subsubsection{Pruning and Refinement}
We employ various quantitative metrics to evaluate the performance and quality of our \method{} model. These metrics include statistical measures such as edge count, node count, edge distribution, and audience distribution. Additionally, we leverage evaluation results generated by the LLMs, incorporating the \textit{Average Edge Score}, representing the mean score of LLM recommendations, and the \textit{Relation Imprecise Rate}, indicating the degree of inaccuracies within the recommendation rationale.

Following the validation results, successive iterations of fine-tuning the prompts for KG generation are undertaken to improve the mentioned metrics. Subsequently, nodes with inferior scores are removed to ensure the quality of the core \method{}. Moreover, erroneous edges, marked by inaccurate or inappropriate relationship descriptions are updated to reflect more reasonable associations between products. This pruning and refinement process serves to augment the overall recommendation quality and contributes to the transparency and interpretability of the \method{}. By iteration refining the graph structure based on validation feedback, we ensure the final \method{} reflects accurate and high-quality recommendation relationships.

\subsection{Internal Product Mapping}

In order to deploy \method{} online, a critical step is to map the LLM output products (nodes in \method{}) to enterprise's internal products. This mapping should address the semantic gap between LLM-recommended products and the actual products available for online recommendation. For generality, we employ a vector search approach; we encode product titles into embeddings with a fine-tuned Bert model, and build them into KNN indexing. Then the mapping becomes a KNN vector search task. It is an efficient and scalable recommendation retrieval approach to capture nuanced semantic relationships between products. The product mapping process involves:

\begin{itemize}
\item \textbf{Generating Product Embeddings.} The initial step involves generating title embeddings for internal enterprise products. Utilizing techniques of Bert embeddings, we transform product titles into dense vector representations. 
\item \textbf{Constructing KNN Index.} Subsequently, we construct a k-nearest neighbors (KNN) index based on these embeddings. This index facilitates rapid retrieval of similar products given a query embedding.
\item \textbf{Mapping Graph Nodes to enterprise Products.} Once the KNN index is constructed, we query the KNN index with the embeddings of knowledge graph nodes, to retrieve the nearest internal products based on semantic similarity. For instance, by querying with the embedding of LLM-generated product title \textit{'Louis Vuitton Damier Graphite Multiple Wallet'}, we can get the nearest product \textit{'Louis Vuitton Damier Graphite Multiple Wallet - Black'} in domain data. By associating (or replacing) each node in the KG with one or multiple internal products, we construct an internal-product-level \method{} for online serving.
\end{itemize}

The fused \method{} encapsulates the essence of the KG while integrating seamlessly with enterprise's internal product inventory, ensuring that we can recommend real products to users.

\subsection{Online Serving} 
We implement an efficient online serving mechanism to leverage the constructed \method{} for real-time recommendation tasks in large-traffic scenarios. We cache every triplet in the KG in a key-value database. Each enterprise product corresponds to a key, with adjacent products and their edges as values. This caching strategy allows for swift access to recommendation information during online serving, offering streamlined retrieval compared to direct searches in a graph database. 

Real-world recommendation systems typically involve two primary stages: candidate selection and ranking. \method{} serves as the candidate selection stage, retrieving a set of products linked to the entry node. Subsequently, the fetched products are further passed to ranker models for more personalized and accurate ranking. By presenting the recommendation to users alongside reasoning, transparency and trust in recommendation process can be fostered.

The \method{} accommodates two types of online recommendation scenarios: item-centric and user-centric. In the item-centric scenario, recommendations are generated based on a seed product, utilizing edges between product nodes. For user-centric scenario, recommendations may originate from user groups, which leverages the edges between products and user groups in the graph. In experiments we focus on the item-centric scenario and leave the user-centric recommendations for furture work.

%% file: experiment.tex
\section{Experiments}
\label{sec:experiment}
In order to evaluate the effectiveness of the \method{} on recommendation performance and user experience, we conducted an online A/B test on an E-commerce website to analyze the performance of our approach in comparison to existing recommendation methodologies.

\begin{table*}[t]
 \centering

 \begin{tabularx}{0.9\linewidth}{c|c|c|p{2.2cm}|p{2.8cm}|c|c}
  \toprule
  \multirow{3}{*}{Metrics} & \multicolumn{4}{c|}{Initial Graph} & \multicolumn{2}{c}{Enterprise Product Graph} \\ 
  \cline{2-7}
  & Nodes & Edges & \small{\textit{Avg Edge Score} (Before pruning)} & \small{\textit{Relation Imprecise Rate} (Before pruning)} & Nodes & Edges \\
  \midrule
   Value & 355,383 & 788,974 & 8.57 & 27.86\% & 325,721 & 718,365   \\ 
  \bottomrule

  \end{tabularx}
 \caption{ \label{tb:statistics} Statistics of the Sneaker \method{} }

\end{table*}

\subsection{Experimental Setup}
\subsubsection{Data Source}
\begin{figure}[ht]
\centering
\includegraphics[width=1\linewidth]{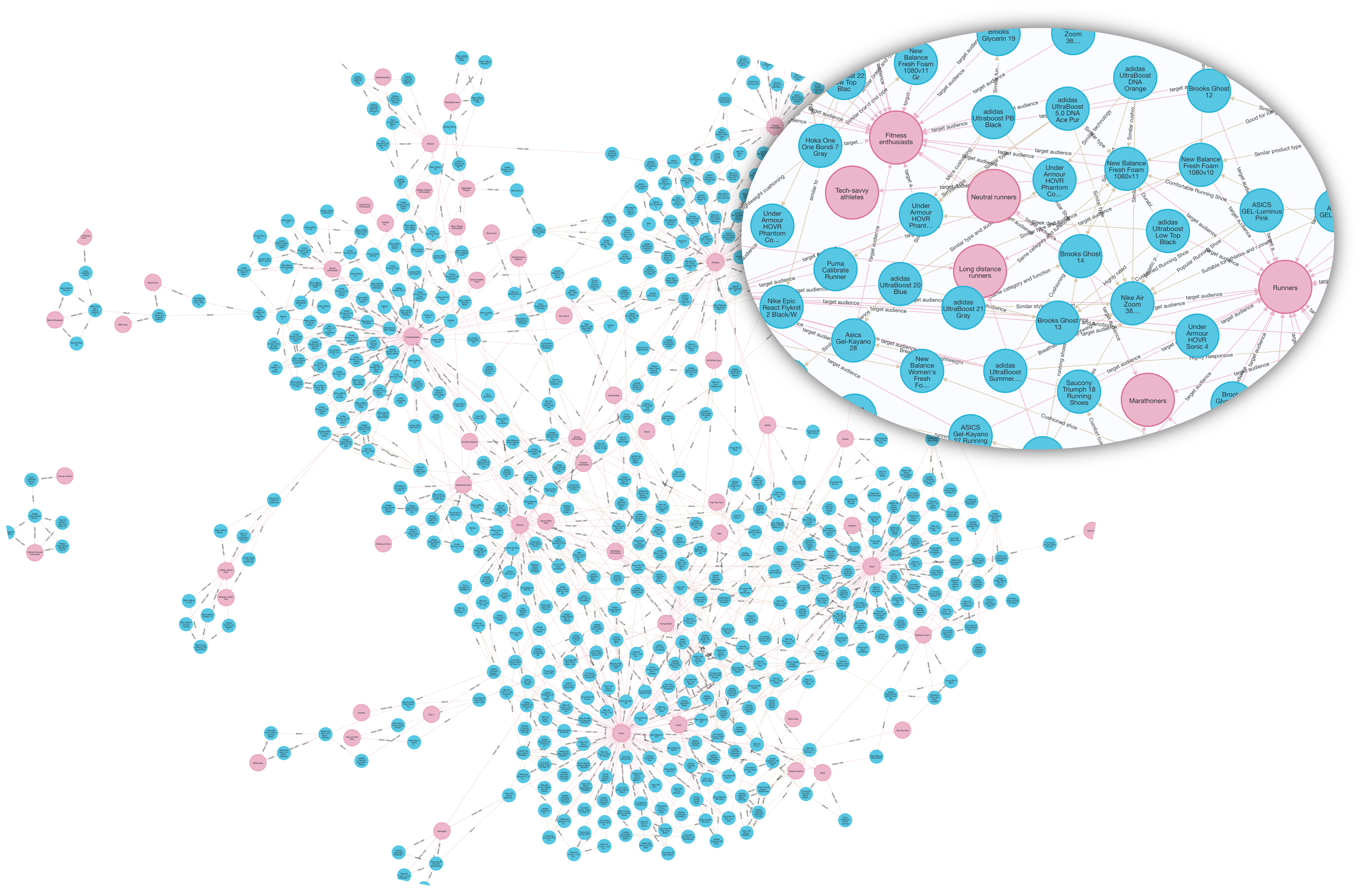}
\caption{\label{fig:coreKGneo4j} Sneaker \method{} Visualization (Sampled)}
\end{figure}

We constructed a sneaker version of \method{}; we curate sneakers from enterprise data and treat them as seed products for KG construction.
Table \ref{tb:statistics} provides a summary of the \method{} statistics. After evaluation and refinement, we removed some nodes and edges. A visualization sample of generated \method{} with Neo4j is provided in Figure \ref{fig:coreKGneo4j}. Due to limitations on the number of nodes that can be visualized effectively, we sampled and displayed 500 nodes from the \method{}. Additionally, we enlarged an area within the visualization, allowing readers to observe the recommendation relationships between nodes and edges.

\subsubsection{Experimental Design}
\begin{figure}[ht]
\centering
\includegraphics[width=1\linewidth]{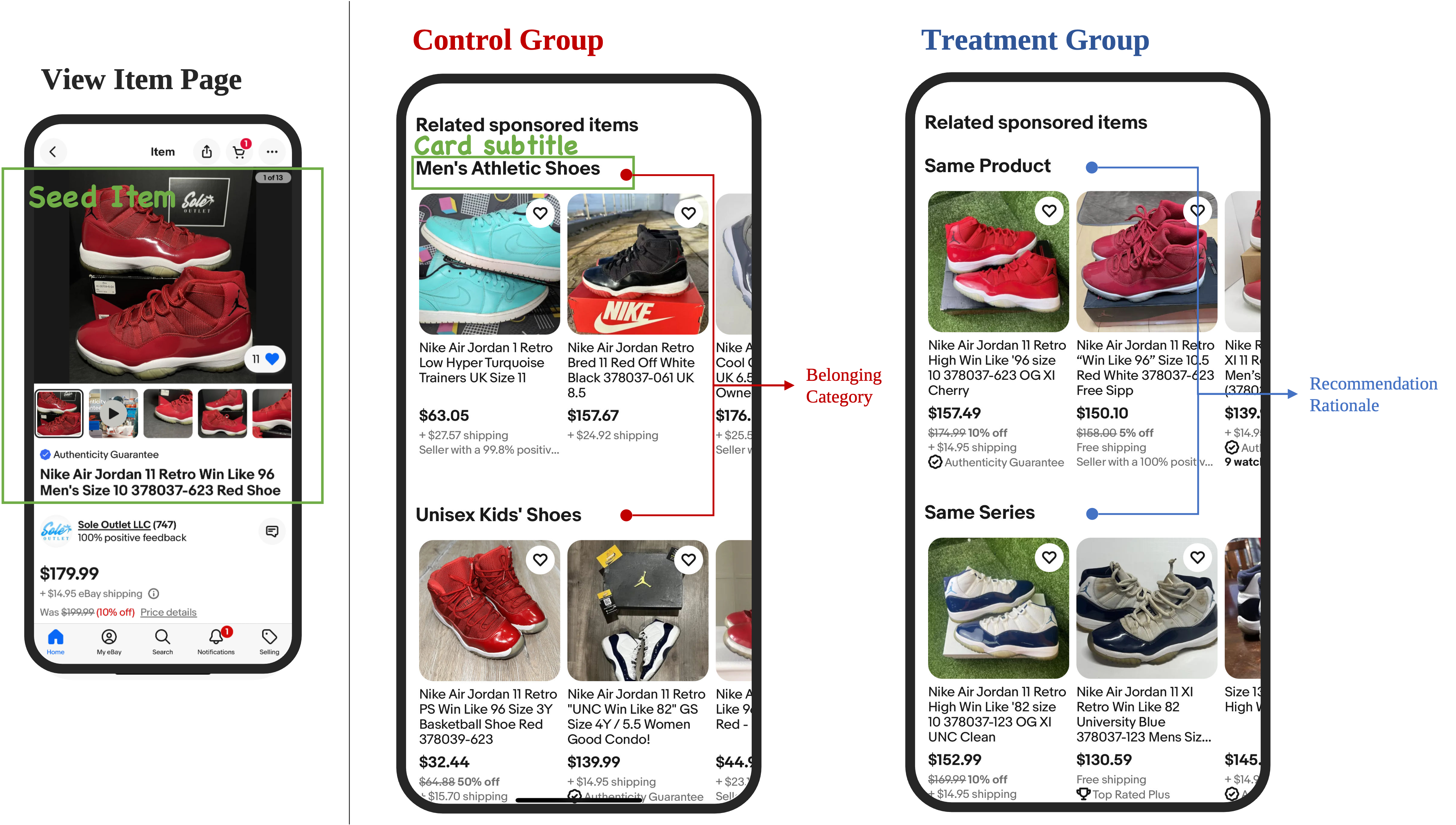}
\caption{\label{fig:experiment} An example of \method{} based recommendation in experiments}
\end{figure}

The A/B test was conducted within the sneaker category, an important domain for the e-commerce platform. The testing placement utilizes a card-group format, consisting of two cards, each displaying 6 item slots within a carousel. Each card can be customized with a subtitle. Positioned on the View Item Page, the test placement appears when users click the \textit{'Add to Watchlist'} button, making it a recommendation placement dependent on the seed item. The illustration of the experiment setup is shown in Fig. \ref{fig:experiment}. 
The primary objective of this A/B test was to compare the performance of two variants: 
\begin{itemize}
\item{} \textbf{Treatment}: The recommended items in the treatment group are recalled by the \method{} and corresponding recommendation rationales are displayed in the subtitle of each card. Additionally, the self-loop edge in the knowledge graph is permitted for recall, retrieving items associated with the same product (here items mean the same product that provided by different sellers). In cases where there are no corresponding nodes in the \method{} for the seed item, the system will fall back to the control. For instance, in Fig. \ref{fig:experiment} we can see subtitles are \textit{'Same Product'} and \textit{'Same Brand'} for two recommendations placements, they are side products of the recommendation based on \method{}.

\item{} \textbf{Control}: The control group adhered to the existing recommendation approach. This group includes recommended items obtained from collaborative filtering and those based on title similarity. In the control group, users were presented with the belonging category names of the recommended items instead of recommendation rationales. For instance, in Fig. \ref{fig:experiment} we can see subtitles are \textit{'Men's Athletic Shoes'} and \textit{'Unisex Kids' Shoes'} for two recommendations placements repsectively, they are fixed texts and may not be always consistent with the items displayed below. 
\end{itemize}

This setup aimed to assess the impact of \method{} on user engagement and satisfaction. Traffic allocation between the Treatment and Control groups was evenly distributed to ensure a balanced and unbiased comparison of their performance. We run the tests with enough data to ensure that the results are statistically significant.

\begin{table}
\centering
  \begin{tabular}{lll}
    \toprule
    Report Metrics & Difference & P-Value \\
    \midrule
    Clicks &+5.19\% & 0.0 \\
    Transactions (1-Day Maturity) & +7.59\% & 0.0007 \\
    Transactions (7-Day Maturity) & +7.55\% & 0.0 \\
    Click Through Rate & +5.91\% & 0.0 \\
    Conversion Rate & +7.20\% & 0.0038 \\
    GMB (1-Day Maturity) & +8.56\% & 0.0459 \\
    Ads Revenue (1-Day Maturity) & +10.84\% & 0.0184 \\
    \bottomrule
  \end{tabular}
\caption{Experimental results}
\label{table:report metrics}
\end{table}

\subsection{Results and Discussion}
The results of the experiment are listed in Table \ref{table:report metrics}. Notable improvements across various metrics were observed in the Treatment group compared to the Control, including Clicks, Transactions, and GMB (Gross Merchandise Bought). 
Specifically, clicks experienced a significant increase of 5.19\% in the Treatment group, indicating enhanced user engagement and interaction with recommended products. Moreover, Transactions and GMB exhibited a substantial positive uplift (7.59\% and 8.56\% respectively), suggesting that users interacting with \method{}-powered recommendations are more likely to initiate transactions compared to those in the Control group. We argue that the benefits of \method{} bring to recommendations are two folds: 1) \method{} can understand the user needs based on their historical behavior, thus providing more accurate recommended items to customers. 2) \method{} can provide explanations of its recommendations to attract users. Unlike other KG based explanation for recommendations, this implementation is very simple in \method{}. We only need to display edge information of KG without any additional operations or modifications to the \method{}, making this approach more efficient and scalable.

In summary, the results of the A/B test show compelling evidence of the efficacy of the \method{} framework in enhancing recommendation performance and user engagement on the website. By providing explainable recommendations accompanied by rationale messages, the \method{} framework effectively satisfies user needs and preferences, leading to increased user interaction, transaction initiation, and overall platform engagement. On the other hand, from the view of knowledge graph, we provide a new method to build knowledge graphs; generate a vanilla KG with LLM first and refine it with domain data. The experimental results also show it is a viable way to apply LLM into web-scale applications.



%% file: conclusion.tex
\section{Conclusion}
In this study, we introduced the novel framework of \method{}, which combines the power of Large Language Models (LLMs) and Knowledge Graphs (KGs) to provide effective and explainable recommendations in the e-commerce domain. 
By leveraging the linguistic capabilities of LLMs and the structured representation of KGs, \method{} enables the item recommendations accompanied by clear rationale messages, enhancing user understanding and trust in the recommendation process. 
Experimental results from the A/B test demonstrated the significant impact of \method{} compared with the control group on driving user engagements and transactions. 
 
Looking ahead, future research directions include further enhancing the \method{} framework by building connections between actual users and user groups within the knowledge graph. Additionally, empirical research on the performance of personalized recommendation by \method{} could provide valuable insights into optimizing recommendation strategies and enhancing user satisfaction.

\begin{table*}
 \caption{Example of LLM Prompts and Response}
  \centering
  \begin{tabular}{p{2.3cm}|p{1cm}|p{12cm}}
    \toprule
    \multirow{30}{*}{KG generation} & Prompt & One user of eBay is browsing a product titled  'Jordan 1 Retro OG High UNC Toe University Blue', hereinafter we refer to it as the seed product.
Provide 5 recommendations that the user might be interested in related to the
seed product. For each recommendation, offer a reasonable rationale in 5
words, then predict the brand, type, and target audience demographics of each
product.

Organize the answer to knowledge graph in JSON format. Nodes should be
products extracted from the seed product and recommendations. Edges should
be (Subject, Predicate, Object) triplets, where predicate is the recommendation rationale from Subject to Object.

An example of output format is as below:
\{"nodes":[\{"product\_title", seed\_product, "brand": "", type: "", audience: ["", "", ""]\}, \{"product\_title": recommendation1\}, ...], 
"edges": [\{"subject":
seed\_product, "predicate", "", "object": recommendation1\}, ...]\}
     \\
    \cline{2-3}
    & Output & \{
  "nodes": [
    \{
      "product\_title": "Jordan 1 Retro OG High UNC Toe University Blue",
      "brand": "Nike",
      "type": "Sneakers",
      "audience": ["Sneakerheads", "Basketball enthusiasts", "Fashion-conscious individuals"]
    \},
    \{
      "product\_title": "Air Jordan 1 Retro High OG 'Chicago'",
      "brand": "Nike",
      "type": "Sneakers",
      "audience": ["Sneakerheads", "Basketball fans", "Streetwear enthusiasts"]
    \},
    \{
      "product\_title": "Nike Air Force 1 Low 'White'",
      "brand": "Nike",
      "type": "Sneakers",
      "audience": ["Streetwear enthusiasts", "Casual sneaker lovers", "Fashion-forward individuals"]
    \},
    \{
      "product\_title": "Adidas Yeezy Boost 350 V2",
      "brand": "Adidas",
      "type": "Sneakers",
      "audience": ["Streetwear enthusiasts", "Fashion-forward individuals"]
    \},
    \{
      "product\_title": "Converse Chuck Taylor All Star",
      "brand": "Converse",
      "type": "Sneakers",
      "audience": ["Classic footwear fans", "Streetwear enthusiasts", "Casual fashion lovers"]
    \},
    \{
      "product\_title": "Off-White x Nike Air Jordan 1 'UNC'",
      "brand": "Nike",
      "type": "Sneakers",
      "audience": ["Hypebeasts", "Fashion-forward sneakerheads", "Streetwear enthusiasts"]
    \}
  ],
  "edges": [
    \{"subject": "Jordan 1 Retro OG High UNC Toe University Blue", "predicate": "Classic colorway appeal", "object": "Air Jordan 1 Retro High OG 'Chicago'"\},
    \{"subject": "Jordan 1 Retro OG High UNC Toe University Blue", "predicate": "Similar brand loyalty", "object": "Nike Air Force 1 Low 'White'"\},
    \{"subject": "Jordan 1 Retro OG High UNC Toe University Blue", "predicate": "Hypebeast appeal", "object": "Adidas Yeezy Boost 350 V2"\},
    \{"subject": "Jordan 1 Retro OG High UNC Toe University Blue", "predicate": "Classic sneaker style", "object": "Converse Chuck Taylor All Star"\},
    \{"subject": "Jordan 1 Retro OG High UNC Toe University Blue", "predicate": "Collaboration hype", "object": "Off-White x Nike Air Jordan 1 'UNC'"\}]\}\\
    \midrule
    \multirow{28}{*}{Edge evaluation} & Prompt & As a e-commerce user browsing a sneaker, you have been provided with recommendations along with accompanying reasons. Your task is to assign a score ranging from 1 to 10 to indicate the acceptability of each recommended item. Additionally, assess the accuracy and appropriateness of the provided reasons. If necessary, provide a better reason. Format all output in JSON format.
    
    User
As an e-commerce user browsing a sneaker, you've received recommendations with accompanying reasons. Your task is to assign a score from 1 to 10 indicating the acceptability of each recommended item. Additionally, evaluate the accuracy and appropriateness of the provided reasons. If needed, suggest better reasons. Please format all output in JSON format.

Here are two examples of input and output :

Input: You are browsing 'Jordan Dub Zero Mid White Cool Grey'. The recommended item is 'Jordan Dub Zero Mid Black White', and the reason is 'Same series'.
    
Output: \{ 
    "Jordan Dub Zero Mid Black White": \{
        "acceptability\_score": 9,
        "reason": \{
            "original": "Same series",
            "accurate": True,
            "alternative": None
        \}
    \}\}

Input: You are browsing 'Jordan Dub Zero Mid White Cool Grey'. The recommended item is 'Jordan Dub Zero Mid Black White', and the reason is 'Same colorway'.

Output: \{ 
    "Jordan Dub Zero Mid Black White": \{
        "acceptability\_score": 9,
        "reason": \{
            "original": "Same colorway",
            "accurate": False,
            "alternative": "different colorway"
        \}
    \}\}

Now you are viewing 'Jordan 1 Retro OG High UNC Toe University Blue'. The recommended item is 'Air Jordan 1 Retro High OG ’Chicago', and the recommendation reason is 'Classic colorway appeal'.
    \\
    \cline{2-3}
     & Output & \{
  "Air Jordan 1 Retro High OG ’Chicago": \{
    "acceptability\_score": 8,
    "reason": \{
      "original": "Same series",
      "accurate": true,
      "alternative": null
    \}
  \}
\} \\
    \bottomrule
  \end{tabular}
  \label{table:example output}
\end{table*}